\documentclass[aip,reprint]{revtex4-1}

\usepackage{graphicx}
\usepackage{amsmath}
\usepackage{braket}
\usepackage{color}
\usepackage{array}
\draft 

\begin{document}

\title{Crossed-Beam slowing to enhance narrow-line Ytterbium Magneto-Optic Traps} 

\author{Benjamin Plotkin-Swing}

\author{Anna Wirth}%
\altaffiliation{Author to whom correspondence should be addressed: \mbox{annaw77@uw.edu}.}

\author{Daniel Gochnauer}

\author{Tahiyat Rahman}

\author{Katherine E. McAlpine}

\author{Subhadeep Gupta}%
\affiliation{Department of Physics, University of Washington, Seattle WA 98195, USA}

\date{\today}

\begin{abstract}
We demonstrate a method to enhance the atom loading rate of a ytterbium (Yb) magneto-optic trap (MOT) operating on the 556 nm ${^1S}_0 \rightarrow {^3P}_1$ intercombination transition (narrow linewidth $\Gamma_g = 2\pi \times 182$ kHz). Following traditional Zeeman slowing of an atomic beam near the 399 nm ${^1S}_0 \rightarrow {^1P}_1$ transition (broad linewidth $\Gamma_p = 2\pi \times 29  $ MHz), two laser beams in a crossed-beam geometry, frequency tuned near the same transition, provide additional slowing immediately prior to the MOT. Using this technique, we observe an improvement by a factor of 6 in the atom loading rate of a narrow-line Yb MOT. The relative simplicity and generality of this approach make it readily adoptable to other experiments involving narrow-line MOTs. We also present a numerical simulation of this two-stage slowing process which shows good agreement with the observed dependence on experimental parameters, and use it to assess potential improvements to the method. 
\end{abstract}


\maketitle 

\section{Introduction} 
Techniques for laser cooling of alkali atoms developed more than three decades ago \cite{metc99} have also been fruitfully applied to laser cooling of atomic species beyond alkalis over the last two decades \cite{bell99,kato99,kuwa99,binn01,grun02,mccl06,lumi10}. New scientific pursuits are afforded through the different electronic structure of such non-alkali atoms including optical atomic clocks \cite{ludl15} for precision metrology and strong dipolar interactions \cite{} for explorations of novel many-body phenomena \cite{ferr18,tang18,trau18}. The different electronic structure can also lead to optical cycling transitions with linewidths far narrower than their alkali counterparts, leading to opportunities for narrow-line laser cooling \cite{} and magneto-optical traps (MOTs) with correspondingly lower temperatures due to the reduced limiting value of the Doppler temperature.

This narrow linewidth however poses a problem for the atom loading rate of a MOT, since the laser cooling force is proportional to this linewidth. One method to circumvent this problem is to use a second transition with a broader linewidth as an intermediate ``pre-cooling'' MOT stage with protection from the correspondingly higher Doppler temperature being furnished by separating the broad- and narrow-line MOT beams in time \cite{kato99,binn01}. 

In the case of alkaline-earth-like ytterbium (Yb) atoms, the standard approach has been to use the broad ${^1S}_0 \rightarrow {^1P}_1$ transition at 399 nm for Zeeman slowing followed by a single color MOT using the ${^1S}_0 \rightarrow {^3P}_1$ transition at 556 nm \cite{kuwa99}. This approach is sufficient for MOT production but typically with loading rate and atom number far lower than in alkali atom MOT setups. Good performance of narrow-line Yb MOTs has been demonstrated by loading it from a 2D MOT operating on the broad transition using a pushing beam \cite{dors13}. In another approach, a broad-line MOT and a narrow-line MOT were simultaneously operated in a core-shell configuration \cite{leej15}. While the core-shell scheme combines the higher loading rate of the broad transition with the lower temperature of the narrow transition, it requires two sets of six MOT laser beams with the additional complexity of shaping these beams to produce a ``shell" arrangement with 399 nm light overlapped with a ``core" of 556 nm light.

In this work we demonstrate a method to enhance the atom loading rate of a narrow-line Yb MOT fed by a Zeeman-slowed atomic beam by introducing only two additional laser beams on the broad transition. These additional beams are oriented in a crossed-beam geometry to provide additional slowing immediately prior to the MOT \cite{footxbeam}. Using this technique, which is readily adoptable to other atomic species, we observe an improvement by a factor of 6 in the MOT loading rate. We also perform a numerical simulation of this two-stage slowing process and find good agreement with the observed dependence on experimental parameters. We then use the simulation to assess potential improvements to the crossed-beam slowing technique.

The rest of this paper is organized as follows. In Section II we discuss the basic idea of the cooling scheme and in Section III present its experimental demonstration. We discuss our numerical model in Section IV and present a summary and outlook in Section V.

\section{Crossed-Beam Slowing}

In order to be captured by a MOT, an atom (i) must be located within the volume of the MOT beams (diameter $D$) and (ii) must be moving slower than the capture velocity $v_c$ of the MOT. A standard method in cold atom physics is to use the Zeeman slower technique \cite{phil82} to reduce the forward velocity of atoms in a beam emerging from an oven and bring a large fraction below $v_c$. This method allows for fine tuning of the exit velocity $v_f$ of the slowed atoms using the current flowing through the electromagnet generating the Zeeman slower field. In traversing the distance $d$ between the end of the Zeeman slower and the MOT beams, the finite transverse velocity $v_t$ of the atoms leads to a transverse displacement $v_t \times d/v_f$ which needs to be less than $D/2$ for capture.     

\begin{figure}[h]
		\center
		\includegraphics[width=0.5\textwidth]{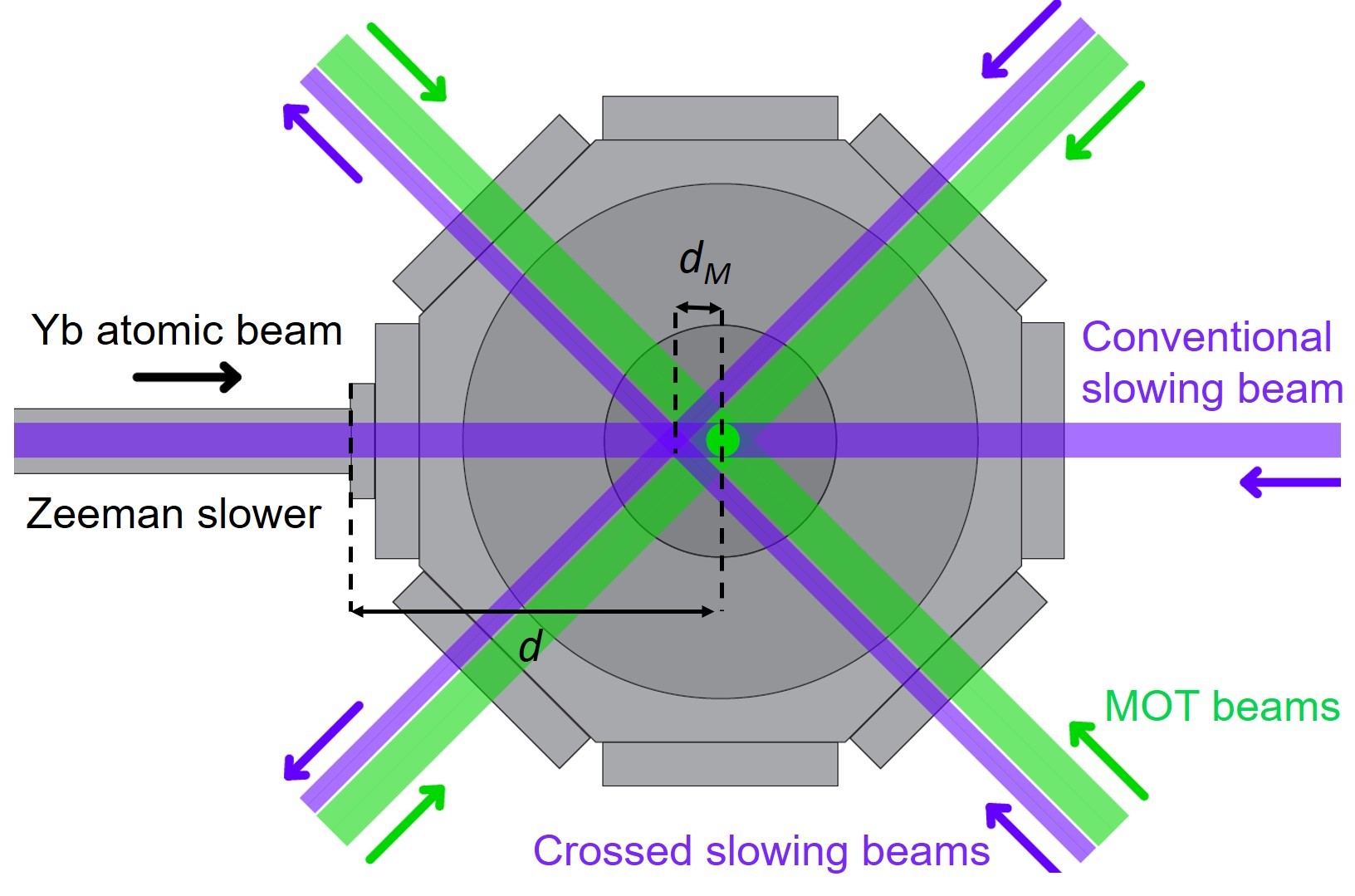}
		\caption{Schematic top-down view of the vacuum chamber with MOT and various slowing beams superimposed. The crossed slowing beams intersect just upstream of the MOT and provide the final stage of slowing down to below the MOT capture velocity. The distances  from the end of the Zeeman slower coils to the MOT center and from the center of the crossed beams to the MOT center are denoted by $d$ and $d_M$, respectively.}
		\label{fig:schematic}
\end{figure}

There are several mechanical and optical access constraints common to such setups (see Fig. \ref{fig:schematic}) that lead to a finite value of $d$. The slowing laser beam is set to counterpropagate with the atomic beam, which means that the beam must pass nearby or through the MOT region. To accommodate this close passage, an increasing field Zeeman slower is usually used so that the slowing beam detuning is sufficiently large to not affect the atoms in the MOT. This scheme requires some distance $d$ for the magnetic field to decay between the end of the Zeeman slower and the MOT region. In addition, the electromagnetic coils at the end of the Zeeman slower are often of sufficient size that they would block optical access to the MOT unless they are at least several inches away. In our experiment $d \simeq 10\,$cm, which is a typical value for such setups. 

Even for an atomic beam that is initially perfectly collimated, transverse velocity at the end of the Zeeman slower will arise from ``blooming" of the atomic beam due to interactions with the slowing beam. This happens because atoms absorb photons from the slowing beam (and receive a momentum kick counter to their motion), and then re-emit these photons in a random direction. Summed over many such events, the average momentum change from emitted photons is zero. However, the distribution of net momentum transfer from emission events has some width, meaning that many atoms do end up with non-zero momentum from emission. Modeling the emission in the transverse direction as a random walk, we expect $v_t$ to scale as $\sqrt{N} v_r$, where $N$ is the number of scattering events in the Zeeman slower and $v_r$ is the recoil velocity. 

We can estimate $v_c \simeq \sqrt{\frac{\hbar k_g \Gamma_g D}{m}}$, where $\Gamma_g = 2\pi \times 182$ kHz is the linewidth of the ${^1S}_0 \rightarrow {^3P}_1$ $556$ nm atomic transition, $k_g$ is the corresponding laser wavenumber, and $m$ is the mass of ytterbium. It is clear from this expression that a narrow-line MOT will feature a correspondingly small capture velocity $v_c$. For Yb this is $9.7\,$m/s for $D=2\,$cm. The transverse velocity is $v_t \simeq 1.5\,$m/s for our system; for an atom traveling at $v_c$, this then converts to a transverse displacement which is larger than $D/2$. These estimates already suggest that loading a narrow-line Yb MOT with a Zeeman slowed atomic beam is less efficient than with a broad-line MOT as in alkali systems. For comparison, in alkali rubidium (Rb) with a large natural linewidth, $v_c$ is almost an order of magnitude larger. See Table I for a summary of characteristics for relevant transitions in Yb as well as Rb.  

The dual constraint on $v_c$ and transverse position on a conventional Zeeman slower is lifted by the addition of a second stage of cooling. The crossed beam slower is designed to provide a final stage of slowing immediately before the MOT, so that the Zeeman slower exit velocity $v_f$ can be set higher than $v_c$. This allows the condition $v_t \times d/v_f < D$ to be satisfied for larger values of $v_t$. The crossed beam slower consists of beams near the strong dipole transition (${^1S}_0 \rightarrow {^1P}_1$ at 399 nm) that are set to propagate parallel to two of the MOT beams, intersecting each other in the atomic beam path just upstream of the MOT, as shown in Fig \ref{fig:schematic}. The transverse forces (vertical in figure) from the crossed slowing beams cancel each other by symmetry, leaving a longitudinal force (horizontal and to the left in figure) that accomplishes the final stage of slowing down to the MOT capture velocity.

\newcolumntype{C}[1]{>{\centering\arraybackslash}p{#1}}
\begin{table}
\caption{\label{tab:table4} Wavelength $\lambda$, linewidth $\Gamma/2\pi$, Doppler temperature $T_{Dop}$, MOT capture velocity $v_c$, recoil velocity $v_{rec}$, recoil shift $\omega_{rec}=\hbar k^2/2m$, and saturation intensity $I_{sat}$ for the two Yb transitions and, for comparison, the main transition in Rb. A MOT diameter of 2 cm is used for the calculation of $v_c$.}
\begin{tabular}{
 C{2.2cm} C{1.7cm} C{1.7cm} C{1.8cm}}
\hline
  & Yb \newline ${^1S}_0 \rightarrow {^3P}_1$ & Yb \newline ${^1S}_0 \rightarrow {^1P}_1$ & Rb \newline ${^2S}_\frac{1}{2} \rightarrow {^2P}_\frac{3}{2}$ \\
  \hline

 $\lambda$ (nm)   & 556    & 399 & 780 \\
 $\Gamma/2\pi$ (MHz)    & 0.182 & 29 & 6 \\
 T$_{Dop}$ ($\mu$K) & 4.4 & 696 & 144 \\
 $v_c$(m/s) &   9.7  & 144 & 67 \\
 $v_{rec}$ (mm/s) &   4.1  & 5.7 & 5.9 \\
 $\omega_{rec}/2\pi$ (kHz) & 3.7 & 7.2 & 3.8 \\
 $I_{sat}$ (W/m$^2$) & 1.38 & 597 & 16.5 \\

\end{tabular}

\end{table}

\section{Crossed-Beam Slower Performance and Characterization}

We demonstrate the performance of the second-stage slower on an apparatus which can produce $^{174}$Yb Bose-Einstein condensates of $10^5$ atoms with cycle times as short as 10 seconds \cite{plot18}. Details of other aspects of the apparatus relevant to cooling Yb can be found in Ref. \onlinecite{plot18thesis}. We now summarize the details relevant for the second-stage slowing. 

The magnetic field profile of the slower consists of an offset field of $B_o \simeq 110\,$G and an increasing field with the total field reaching a maximum value of $B_f \simeq 475\,$G at the exit of the slower. The maximum initial forward velocity that can get slowed by the slower is then $\mu_B (B_f-B_o)/(\hbar k_p) \simeq 200\,$m/s, where $\mu_B$ is the Bohr magneton and $k_p$ is the laser wavenumber of the ${^1S}_0 \rightarrow {^1P}_1$ 399 nm transition. In practice, atoms starting with even higher forward velocity from the oven can be slowed by our apparatus due to the slowing laser beam also interacting with the atoms in the space between the oven and the start of the Zeeman slower. $B_f$ is adjustable with the current supplied to the electromagnet generating the increasing field. The slowing laser beam addresses the 399 nm transition and is detuned by $\delta = - 2 \pi \times 807$\,MHz $(\simeq -\mu_B B_f/\hbar)$ from the transition with an average intensity approximately equal to the saturation intensity.

The two crossed laser beams forming the second-stage slowing (see Fig.\ref{fig:schematic}) are positioned to intersect at a distance $d = 10$ cm beyond the end of the Zeeman slower coils and $d_M = 1$ cm before the MOT. The crossed beams are elliptical, with the long axis oriented in and out of the page with respect to Fig. 1, and with the short axis having an approximately Gaussian horizontal profile of $1/e^2$ width 1.5 mm. The dimension of the ellipse long axis is set to make the height of the crossed beam slowing region approximately match the diameter of the MOT beams. The beam is made narrower on the other axis so that the crossed region could be placed as close to the MOT as possible  without disturbing the atoms trapped in the MOT. The frequency of the crossed beams was experimentally optimized to be $\delta_X=-2\pi \times 42$ MHz detuned from the 399 nm transition. 

\begin{figure}[h]
		\center
		\includegraphics[width=0.5\textwidth]{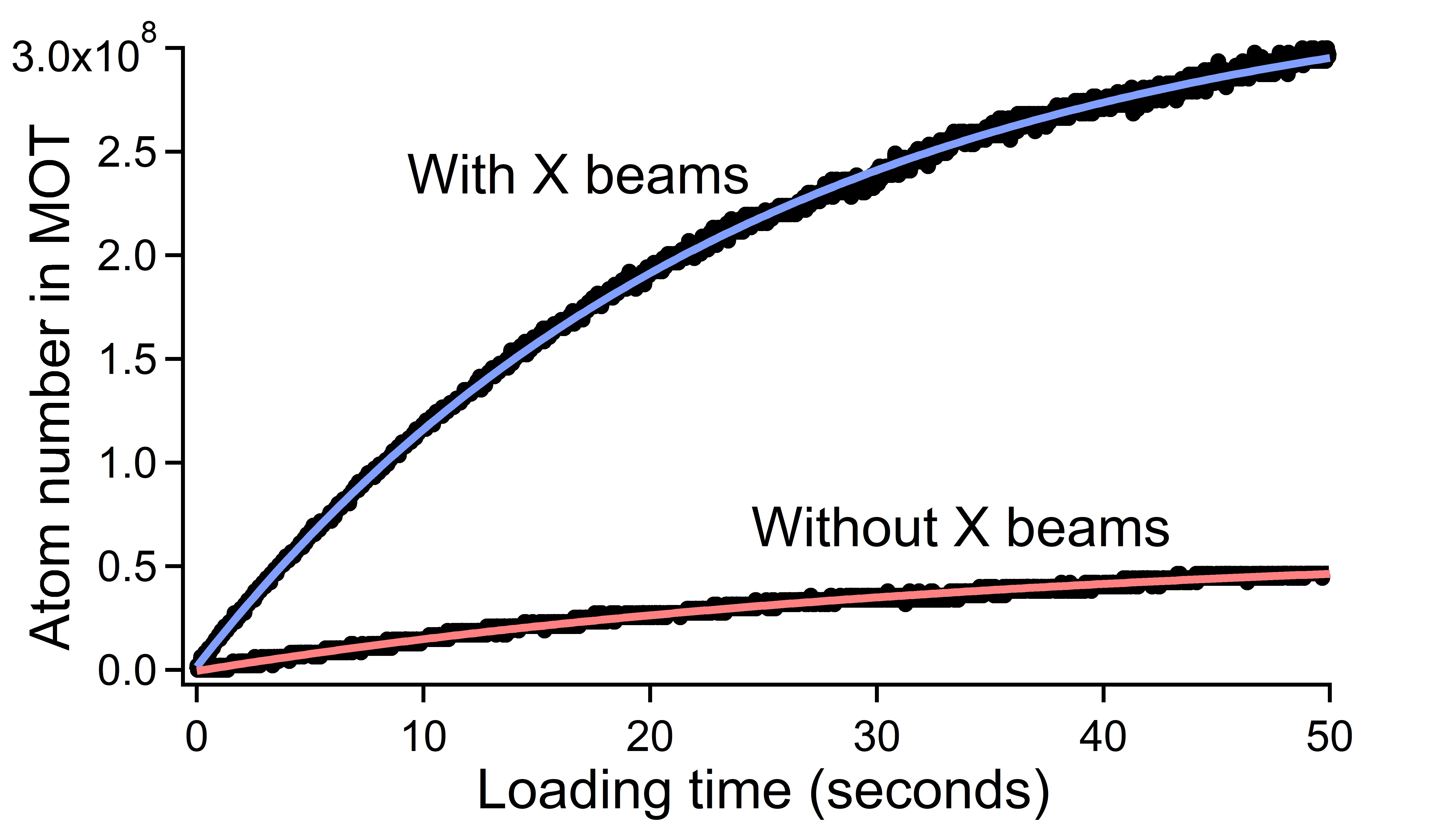}
		\caption{Fluorescence signals (black lines) showing the atom number growth in a
        narrow-line Yb MOT for optimized arrangements with and without the
        crossed-beams. For each of the two data curves, the slower current was
        adjusted to maximize the loading rate and $s_X = 0.3$ for the with-crossed-beams data. The data are  fit to exponential curves for both
        with crossed beams (blue line) and without crossed beams (red line)
        cases. The loading rates are given by the initial slopes which are
        different by a factor of 6.}
		\label{fig:loadcurves}
\end{figure}

We assessed the performance of the crossed-beam slower by comparing the MOT loading rates for various parameters of crossed-beams and slowing beam. Representative ``loading curves" are shown in Fig. \ref{fig:loadcurves}. Such loading curves were obtained by monitoring the fluorescence of the MOT using a photomultipler tube and are fitted by a function of the form $N(t)=N_0(1-e^{-Lt/N_0})$, where $L$ is the initial atom loading rate and $N_0$ is the equilibrium number at long times. When utilizing the crossed beams, we see a marked improvement in both the MOT loading rate and overall MOT population. 

\begin{figure*}
		\includegraphics[width=\linewidth]{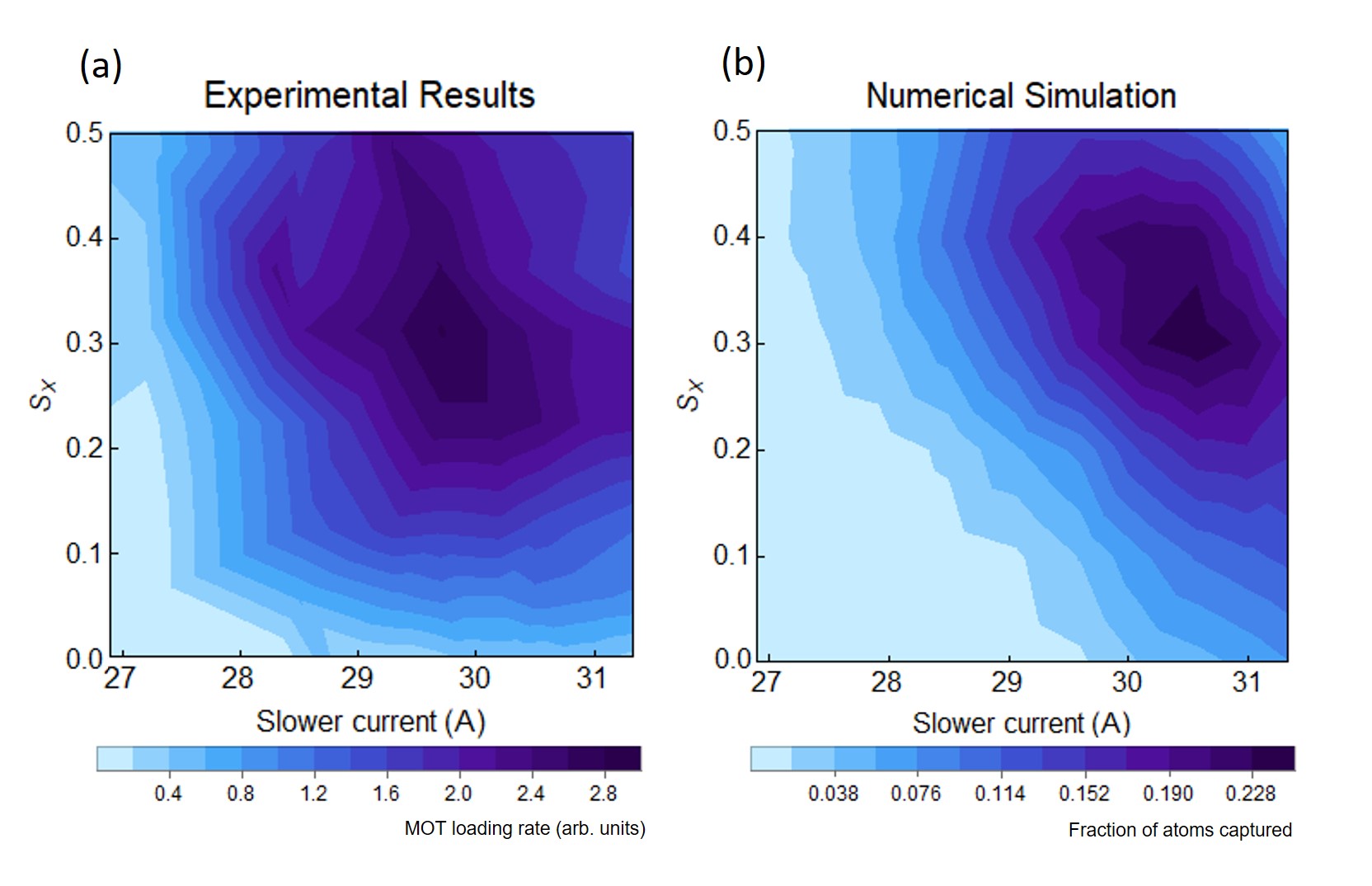}
		\caption{Contour map of the MOT loading rate versus saturation intensity of the crossed beams and Zeeman slower current. Color indicates the relative fraction of atoms captured in the MOT. The optimal atom capture occurs for a non-zero crossed beam intensity. (a) MOT loading rate from fluorescence measurements  at various Zeeman slower currents and crossed beam intensities. Map is built up of 64 data points. (b) Simulated results for the fraction of atoms exiting the Zeeman slower which are captured by the MOT. Map is built up of 66 data points.}
		\label{fig:heatmaps}
\end{figure*}

To further explore the performance of the crossed beam slower, we mapped out the behavior of the loading rate in the two-dimensional parameter space of slower current and crossed beam intensity $s_X$ (see Fig. \ref{fig:heatmaps}(a)). The success of the technique is gauged by observing that the peak occurs at a finite value of $s_X$ and that the loading rate at the peak is significantly greater (by a factor of about 6) than the largest loading rate along the $s_X=0$ axis. Furthermore we see that the location of the largest loading rate for a given $s_X$ moves towards lower currents as $s_X$ is increased. This is expected because by increasing the slowing power of the crossed beams, the atoms may have a higher velocity coming out of the Zeeman slower and still be captured by the MOT. 

\section{Numerical Model}
To provide a theoretical model for our experimental results, we numerically simulate the trajectories of atoms subject to laser cooling forces through the experimental apparatus. After verifying that it captures the main features of our experimental results, we also use this model to analyze prospects to improve the cooling performance.

\subsection{Description of Model and Comparison to Experiment}
Within the Zeeman slower, an atom experiences a position- and velocity-dependent average scattering force from the slowing beam with acceleration given by:
\begin{equation}
a_S(v,z)\!=\!\frac{-\hbar k_p \Gamma_p}{2m}\left[\frac{s}{1\!+\!s\!+\!\frac{4}{\Gamma_p^2}\!\left(\delta\!+\!k_pv\!+\!\frac{\mu_B B(z)}{\hbar}\right)^2}\right]
\label{eq:slower}
\end{equation}
where $B(z)$ is the magnetic field $z$ along the longitudinal direction \cite{plot18thesis} and $v$ is also a function of $z$.

In the region of the crossed beams, the slowing beam is far off resonance with the atoms that are moving slow enough to be eventually captured by the MOT. The average scattering force from the two-crossed beams is:
\begin{equation}
a_X(v,z)\!=\!\frac{-\hbar k_p \Gamma_p}{2m}\left[\frac{\frac{2}{\sqrt{2}} s_X(z)}{1\!+\!s_X(z)\!+\!\frac{4}{\Gamma_p^2}\left(\delta_X\!+\!k_p\frac{v}{\sqrt{2}}\right)^2}\right]
\end{equation}
where $s_X(z)$ is the saturation parameter of the crossed beams at position $z$. The factor of $2/\sqrt{2}$ in the numerator accounts for the horizontal components (to the left in Fig. \ref{fig:schematic}) of the force from the two crossed beams which intersect the atomic beam each at an angle of 45 degrees. The force components in the orthogonal direction (up/down in Fig. \ref{fig:schematic}) cancel each other out. The factor of $1/\sqrt{2}$ in the Doppler term in the denominator is also from the crossed beams propagating at 45 degrees with respect to the atomic beam. The spatial dependence of $s_X$ is from the Gaussian transverse profile of the beams. 

\begin{figure}
		\includegraphics[width=\linewidth]{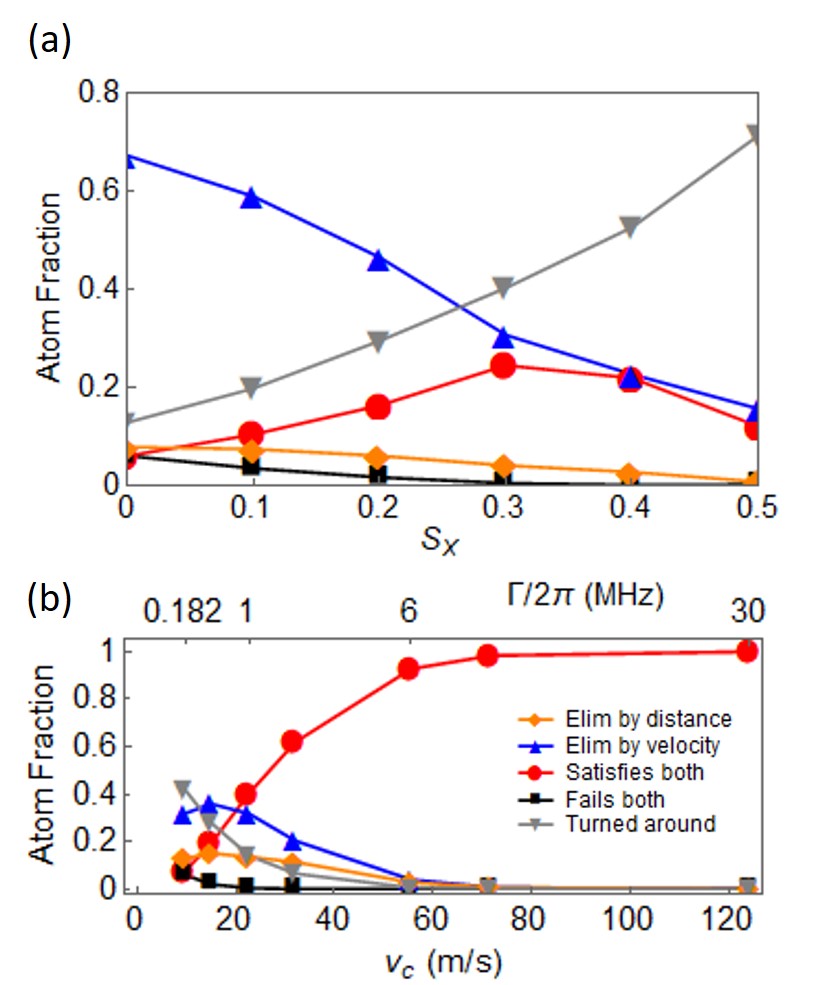}
		\caption{Numerically simulated trajectories of slowed atoms categorized into five fractions - those that fail the distance condition (i) only (yellow diamonds); those that fail the velocity condition (ii) only (blue triangles pointing up); those that satisfy both (i) and (ii) and are captured by the MOT (red circles); those that fail both (i) and (ii) but continue to move forward (black squares); and those that have their longitudinal velocities reversed by the slowing forces (gray triangles pointing down). (a) shows these fractions for different crossed-beam intensities and fixed slower current of 30.7 A (b) shows these fractions for different values of the MOT capture velocity $v_c$ and with $s_X=0$. For each point in (b), the slower current is adjusted to optimize the captured fraction.}
		\label{fig:atomfractions}
\end{figure}

We first simulate the trajectory of atoms through the Zeeman slower using Eq. (1). This provides a conversion between slower current and the atom longitudinal velocity $v_{center}$ entering the crossed-beam region. In practice, the distribution of $v_{center}$ is broadened by the natural linewidth $\Gamma_p$ of the transition and non-ideal effects such as the finite laser linewidth, irregularities in the slowing beam profile, and fluctuations in current to the Zeeman slower coils. We approximate these effects in our simulation using a Gaussian distribution with standard deviation $12\,$m/s, as measured in earlier Doppler spectroscopy characterizing the Zeeman slower in our apparatus \cite{plot18thesis}, and consistent with $\Gamma_p/k_p$ for this transition.

For a given $s_X$ and slower current, we then simulate the trajectories of 3000 atoms through the crossed-beams with initial longitudinal velocities randomly selected from a normal distribution with mean $v_{center}$ and standard deviation 12 m/s. For both Eqs (1) and (2), we use a time increment of $20\,\mu$s for the simulations. 
 
To model the effect of spontaneous emission, the calculated number of scattered photons is used as the number of steps and $v_{rec}$ as the step size in a 3-dimensional random walk, which increases the transverse velocity $v_t$. Upon reaching the MOT beams, an atom is captured if it simultaneously satisfies two conditions: \\
(i) its transverse distance from the center is less than the $1\,$cm radius of the MOT beams and \\
(ii) its total final velocity $v_{tot}= \sqrt{v_{l}^2+v_{t}^2}$, where $v_l$ is the longitudinal velocity, is less than $v_c=9.7\,$m/s. 

Fig. \ref{fig:heatmaps} shows how these simulations compare to our experimental data. We see that optimal capture for our experimental parameters occurs at around slower current of $30\,$A and $s_X=0.3$. The simulations demonstrate good agreement with the experimental results, with the peak locations differing by less than 3\% of the slower current. 

\subsection{Analysis for Optimization of Crossed-Beam Slowing}

Since the numerical simulation is in agreement with the experimental data, we now use it for further assessment of the crossed-beam slowing method in order to assess potential improvements. We categorize the trajectories of the Zeeman-slowed atoms within the crossed-beams into five possibilities - those that fail the distance condition (i) only; those that fail the velocity condition (ii) only; those that satisfy both (i) and (ii) and are captured by the MOT; those that fail both (i) and (ii) but continue to move forward; and those that have their longitudinal velocities reversed by the slowing forces. 

The impact of the crossed beams is shown in Fig. \ref{fig:atomfractions}(a) where the calculated distributions across the five categories are plotted for different crossed-beam intensities at the optimum Zeeman slower current of 30.7 A. We emphasize that we are only considering the slowed flux emerging from the Zeeman slower and ignoring the un-slowed atoms that are out of consideration for MOT capture. As $s_X$ increases, atoms that went through the MOT beams but were too fast to be captured are slowed down by the crossed beams (decay of blue curve) and can now be captured (growth of red curve). Increasing $s_X$ also leads to more atoms being turned around (gray curve) and an optimum is reached at $s_X=0.3$ when the red curve peaks. Importantly, the atom fraction not captured due to either too low (yellow) or negative (gray) velocities is only $20\%$ at $s_X=0$. In comparison, this value is far greater for a MOT optimized without the crossed-beams, and in our case is about $55\%$ (this situation is depicted by the first set of points in Fig. \ref{fig:atomfractions}(b) for $v_c=9.7$ m/s, discussed below). This demonstrates how the crossed-beam method counters the blooming problem.    

For our experimental parameters, the optimum captured fraction with the aid of the crossed-beams is about 25$\%$ of the slowed atom flux emerging from the Zeeman slower. As we have shown, this is significantly better than the performance without the crossed-beams by solving the blooming issue. To further support this point, we have also used our numerical model to verify that reducing the distance $d$ between the end of the slower and the MOT provides a negligibly small improvement for the optimum captured fraction of $25\%$. The largest remaining factor that limits our captured fraction comes from the width 12 m/s of $v_{center}$, which is larger than the MOT capture velocity $v_c$ of 9.7 m/s. Using our model, we have also simulated the effect of having a larger $v_c$ which in turn can be produced by either larger diameter MOT beams or a larger transition linewidth. The improvement to captured fraction is then expected even without the crossed beams, as shown in Fig. \ref{fig:atomfractions}(b), where the captured fraction rises from around 5$\%$ to essentially 100$\%$ corresponding to linewidth rising from $\Gamma_g$ to $\Gamma_p$. 

For a core-shell MOT \cite{leej15} we would expect a maximum $v_c$ corresponding to $\Gamma_p$. Since the crossed-beam slowing method achieves a peak captured fraction of $25\%$, this suggests that for our current experimental parameters, the crossed-beam method captures about 4 times less than the core-shell MOT method. However, the crossed-beam method is a technically simpler alternative since it only requires two beams instead of six additional MOT beams and it also avoids the additional complexity of shaping each of these beams to produce a ``shell" arrangement with 399 nm light overlapped with a ``core" of 556 nm light. Furthermore, compared to Ref. \onlinecite{leej15}, the total power requirements are a factor of four less, making the crossed-beam setup more cost-effective. 

As suggested by Fig. \ref{fig:atomfractions}(b), further improvement in the crossed-beam method is possible for larger MOT beam diameter providing a larger $v_c$. Using our numerical model we have calculated that at $s_X=0.3$, we can expect a captured fraction of 40$\%$ for a beam diameter that is twice larger than our current value, making the performance comparable to that of the core-shell MOT.

\section{Summary and Outlook}
We have described a method to improve the performance of a narrow-line Yb MOT, demonstrating a factor of 6 enhancement, using a crossed arrangement of two additional slowing beams immediately prior to the MOT. We have assessed the performance of our method over a large parameter space of slower magnetic field and crossed beam intensity. Using a numerical model for our system that shows good agreement with our experimental observations, we have also demonstrated the efficacy of this method to counter the blooming problem, with future improvements possible by increasing the size of the MOT beams. Compared to the recently demonstrated core-shell MOT method \cite{leej15}, our analysis indicates that our performance with current parameters is about a factor of four lower, with improvements to within a factor of about two expected with larger MOT beams. Importantly, our method is substantially simpler to technically implement, involving fewer laser beams and far lower power requirements. 

Our results can be adapted to other experimental efforts which use laser cooled Yb, an atom with various applications in atomic clocks \cite{hink13}, preparation of quantum degenerate systems \cite{taka03,hans11}, precision atom interferometry \cite{jami14}, quantum simulation \cite{paga14,scaz14}, and quantum information processing \cite{stoc08,dale08}. The method is also applicable to narrow-line MOTs of other elements and has been very recently demonstrated in Dy \cite{lund19} and Er \cite{seob19}, where the linewidths of the broad and narrow transitions are similar to those for Yb.  
\section*{Acknowledgments}

We thank Alan Jamison for useful discussions. This work was supported by NSF Grant No. PHY-1707575.   

\section*{Data Availability}
The data that support the findings of this study are available from the corresponding author upon reasonable request.

\bibliographystyle{apsrev}
\bibliography{ifmrefs20}

\end{document}